\documentclass[]{fairmeta}

\usepackage[english]{babel}

\usepackage{amsmath}
\usepackage{amsfonts}
\usepackage{graphicx}

\title{From Thought to Action: How a Hierarchy of Neural Dynamics Supports Language Production}

\author[2,3]{Mingfang (Lucy) Zhang}
\author[1]{Jarod Lévy}
\author[1]{Stéphane d'Ascoli}
\author[1]{Jérémy Rapin}
\author[4]{F.-Xavier Alario}
\author[2, 5]{Pierre Bourdillon}
\author[6,7*]{Svetlana Pinet}
\author[1*]{Jean-R\'emi King}

\affiliation[1]{Meta AI}
\affiliation[2]{Hospital Foundation Adolphe de Rothschild}
\affiliation[3]{École Normale Supérieure, Université PSL, CNRS}
\affiliation[4]{Aix Marseille Université, CNRS, CRPN}
\affiliation[5]{Paris Cité University}
\affiliation[6]{Basque Center on Cognition, Brain and Language, San Sebastián}
\affiliation[7]{Ikerbasque, Basque Foundation for Science, Bilbao}

\contribution[*]{Equal contribution}

\abstract{
Humans effortlessly communicate their thoughts through intricate sequences of motor actions. Yet, the neural processes that coordinate language production remain largely unknown, in part because speech artifacts limit the use of neuroimaging. To elucidate the unfolding of language production in the brain, we investigate with magnetoencephalography (MEG) and electroencephalography (EEG) the neurophysiological activity of 35 skilled typists, while they typed sentences on a keyboard. This approach confirms the hierarchical predictions of linguistic theories: the neural activity preceding the production of each word is marked by the sequential rise and fall of context-, word-, syllable-, and letter-level representations. Remarkably, each of these neural representations is maintained over long time periods within each level of the language hierarchy. This phenomenon results in a superposition of successive representations that is supported by a hierarchy of dynamic neural codes. Overall, these findings provide a precise computational breakdown of the neural dynamics that coordinate the production of language in the human brain.
}

\date{\today}
\correspondence{
\email{lucy.zhang@psl.eu}, 
\email{s.pinet@bcbl.eu},
\email{jeanremi@meta.com}}

\begin{document}
\maketitle

\section{Introduction}

\begin{figure}[t!]
    \centering
    \includegraphics[width=\linewidth]{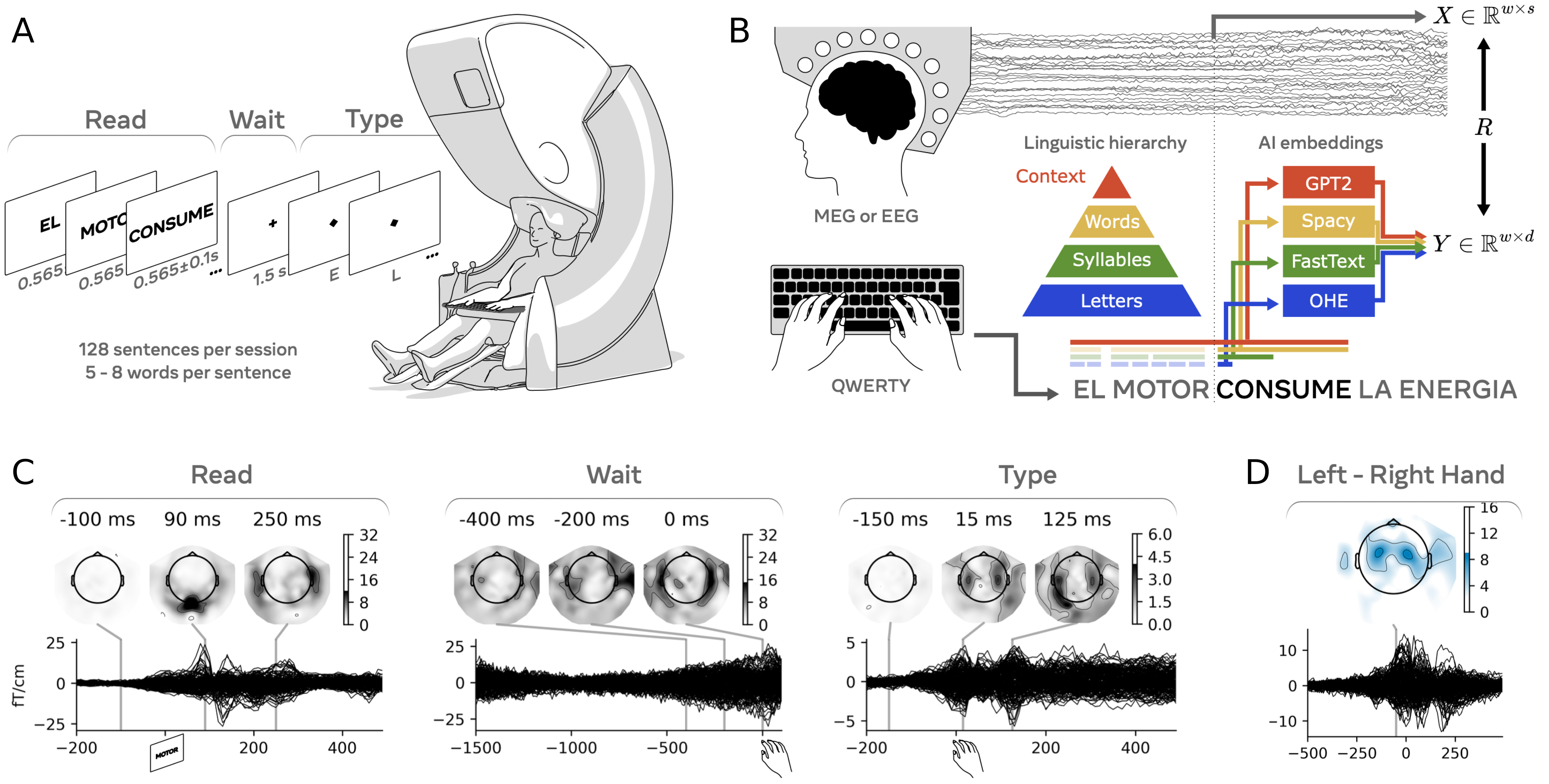}
    \caption{\textbf{Experimental design and methods.} \textbf{A}. We recorded the brain activity of healthy volunteers with either MEG or EEG during a read-wait-type task, in which subjects typed briefly memorized sentences. 
    \textbf{B.} To investigate when and whether the brain produces a hierarchy of language representations, we linearly decoded, from these signals ($X$), the vectorial embeddings ($Y$) of four levels of representations: contextual word embeddings (with GPT-2), de-contextualized word embeddings (with Spacy), syllable embedding (with FastText) and letters (with a One-Hot-Encoder, OHE), and summarize the decoding scores with a Pearson correlation ($R$). \textbf{C.} Median MEG responses time-locked to word presentation (left), first keystroke (middle), and all keystrokes (right). 
    \textbf{D.} Difference in MEG responses between keys typed with left vs right hands. A deep learning pipeline optimized to decode sentences from these data is available in our companion paper \citep{levy_brain--text_2025}.}
    \label{fig1}
\end{figure}

Humans seamlessly produce intricate sequences of motor actions to communicate their thoughts. 
Such language production ability systematically requires a complex chaining of motor actions, whose underlying neural mechanisms remain poorly understood. 

An extensive cortical network is involved in the production of speech \citep{indefrey_spatial_2011, price_review_2012}, typing \citep{pinet_tracking_2019, pinet_response_2016,hu_precision_2023}, writing \citep{planton_handwriting_2013, cloutman_neural_2009}, and signing \citep{blanco-elorrieta_language_2018, emmorey_neural_2007}. 
The associative cortices of the temporal and parietal regions primarily build and maintain linguistic features \citep{matchin_functional_2022, fedorenko_language_2024, chang_information_2022} in interaction with inferior frontal and supplementary motor areas \citep{bourguignon_rostro-caudal_2014}, while also projecting to pre-motor and motor cortices for action planning and execution \citep{hickok_computational_2012, guenther_neural_2016}. 
This network is thought to enable language production through a hierarchical process. According to this view, the representation of a sentence meaning would trigger an incremental retrieval and encoding of words, syllables, and letters or phonemes \citep{bock_language_1994,dell_language_1997, levelt_word_2000, logan_hierarchical_2011, roelofs_weaver_1997, rumelhart_simulating_1982}.

However, two main challenges have hindered the characterization of the timing, content, and coordination of the neural bases of natural language production.
First, speech generates auditory feedback, which complicates the discrimination of perception and production processes, and has thus led to using artificial paradigms such as single word production and covert speech tasks \citep{Salmelin1994a, carota_time_2022}.
Second, speech production generates facial artifacts in functional Magnetic Resonance Imaging (fMRI), electro- (EEG) and magneto-encephalography (MEG) \citep{abbasi_correcting_2021, volfart_comparison_2024, vos_removal_2010}. While this issue can be  circumvented through the artifact-free yet sparsely-sampled intracranial recordings of epileptic patients, current methods do not readily allow a comprehensive whole-brain study of natural language production.

To address these challenges, we evaluate here the neural bases of natural language production during typing.
This approach presents two major advantages compared to speaking: (i) most of its sensory feedback (the letters on the screen) can be removed and (ii) muscle activity remains distant from M/EEG sensors. 
Thus, we recorded, with either MEG or EEG, the brain activity of 35 skilled typists while they typed briefly memorized sentences (Fig. \ref{fig1}A).
To characterize the hierarchy of language representations generated in the brain during production, we rely on the classic definition of ``neural representations'' as information linearly decodable from neural activity \citep{dicarlo_untangling_2007}. 
To study language representations in the brain, we featurize letters, syllables, words, and contexts into simple vector representations or those from pre-trained language models (Fig. \ref{fig1}B). 
Finally, we evaluate, with linear decoding, whether and when these hierarchical representations rise and fall during sentence production, and summarize these decoding scores with either accuracy or a Pearson correlation (R).
For clarity, and unless stated otherwise, we here focus on the MEG results. We direct the reader to the supplementary materials for the replication of these analyses on EEG data (see \Cref{Supplementary}). 
To explore the possibility of efficiently decoding text from these brain signals, we refer the reader to our companion paper \citep{levy_brain--text_2025}. 

\section{Results}

\paragraph{Task-related neural responses.}
Before decoding language representations from MEG signals, we first examined the average brain responses in the read, wait, and type phases in this task (\Cref{fig1}A).
During the reading phase, word presentation elicits a posterior MEG response between 100 and 200\,ms after their onset, followed by a bilateral and more anterior activation (\Cref{fig1}C, see EEG in \Cref{eeg topo}). %
Next, the wait period is marked by an increase in activation in the lateral MEG channels. 
Third, the brain response time-locked to keystrokes is marked by a bilateral motor topography, followed by a peak in the left-lateralized sensors. A similar effect can also be seen by comparing left- vs right-hand keystrokes (\Cref{fig1}D).
Overall, these topographies are consistent with those found during language processing (reading network), the preparation of motor actions, and their actual execution \citep{munding_cortical_2016}. 
\begin{figure}[t!]
    \centering
    \includegraphics[width=\linewidth]{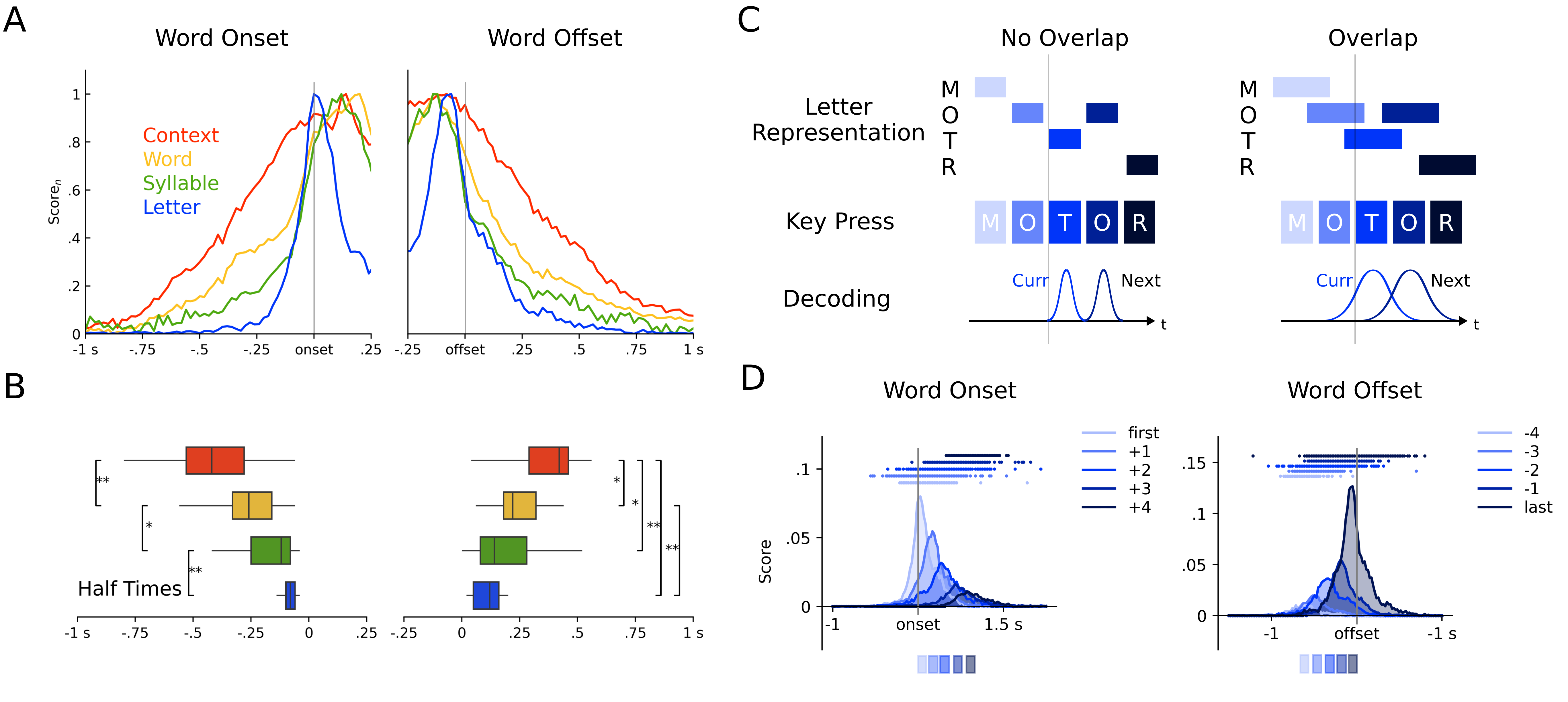}
    \caption{\textbf{Decoding a hierarchy of linguistic features.} \textbf{A.} Normalized decoding score of each linguistic feature time-locked to word onset and offset. \textbf{B.} Box plot of half-times (when the decoding reaches half of the maximum score) across participants. Asterisks, * and ** indicate significant ($p<0.05$ and $p<0.01$, respectively) between the half times of different representations. For clarity, we show skip-level comparisons only if adjacent-level tests are not consistently significant. 
    \textbf{C.} Two hypotheses (left/right column) for the neural representations of successive letters. 
    Top: each row corresponds to a putative neural activation of one letter representation.
    Middle: typing behavior.
    Bottom: putative decoding scores. 
    \textbf{D.} Shifted time-decoding of successive letters relative to word onset or offset. Each curve corresponds to the decoding score at a letter position. 
    Top lines indicate significant decoding scores ($p<0.05$). Each rectangle below indicates the median start and duration of the corresponding key. 
}
    \label{fig2}
\end{figure}

\paragraph{Isolating the neural representations of individual key presses.}
Current theories predict that language production depends on a hierarchy of language representations.
We thus hypothesize that: 
(1) different levels of representations will be linearly decodable from brain activity, and
(2) higher-level representations will be detected before lower-level representations.


To test these hypotheses, we first evaluated whether each letter can be linearly decoded from the MEG activity relative to words.
For this, we fitted a series of linear classifiers on all MEG channels at a single time sample relative to word onset and evaluated them on held-out MEG responses. 
We repeated this process for word-offset responses. 
Due to the class imbalance induced by the inherent differences in letter frequency, it can be difficult to interpret the chance level performance. For clarity, we thus report the difference in balanced accuracy between this linear decoder and a dummy classifier (chance level of 0\%, details see \Cref{Method}). 
Decoding of the first letter of each word peaks around the corresponding key press with a performance of 8.1\% (SEM across subjects: $\pm$0.6\%), but is already significant from -220\,ms (\Cref{meg raw score sup}).
Qualitatively similar decoding is observed with EEG, although with a much lower decoding performance: Score=1.6\% (SEM across subjects: $\pm$0.5\%)  (\Cref{eeg sup hierar}).

\paragraph{Tracking higher-level representations of language.}
To extend this approach to higher linguistic features, we adapted this linear decoding analysis to syllables. 
However, the 288 distinct syllables present in this dataset make classification intractable. 
We thus embedded each syllable in FastText \citep{bojanowski_enriching_2017}, a character-level deep neural network trained on language modeling. This model transforms each syllable into a 300-dimensional vector, whose geometry represents the orthographic regularities of large text corpora. For clarity, we here report the decoding of the first syllable of each word relative to word onset. 
The resulting decoding shows that these syllable representations can be detected from brain activity up to -520\,ms before word onset,
and ultimately peak at +120\,ms with decoding of R=0.10$\pm$0.01. The corresponding EEG decoding can be found in \Cref{eeg sup hierar}. 

\paragraph{Decomposing the hierarchy of linguistic features.}
We generalized this embedding approach to two additional levels of the language hierarchy. Word embedding, as provided by Spacy \citep{ines_montani_explosionspacy_2023}, represents words as 300-dimensional vectors such that words used in the same contexts are maximally similar. Contextual word embedding, as extracted from GPT2 \citep{radford_language_2019}, follows a similar principle for the \emph{combination} of words and can thus embed context. 
To get causal representations, we here focused on the context up to the current word. 
Linear decoding shows that both of these levels of representations are represented in brain activity long before the corresponding word is typed. Decoding of word representations peaks with a R=0.17$\pm$0.01\%, while decoding of context embedding peaks with a R=0.13$\pm$0.01\%. The same results for EEG can be found in \Cref{eeg sup hierar}. 
%

\paragraph{A top-down sequence of activations precedes word production.}
Theories and empirical findings consistently highlight that language production is supported by a hierarchy of representations. 
To explore the unfolding of this process, we evaluated whether the rise of each representation depends on its level in the language hierarchy.
For this, we estimated, for each subject, the ``half-time'' of each neural representation -- the time point where decoding reaches 50\% of its maximal value within the window of focus. 
The Spearman correlation between half-times and linguistic levels ordered from context to letter is significant across subjects: R=-0.77$\pm$0.06, $p<10^{-5}$. 
A similar, although noisier, finding is also replicated with EEG: R=-0.58$\pm$0.07, $p<10^{-4}$ (\Cref{eeg sup hierar}). 
The brain thus appears to produce language in a hierarchical fashion, by first generating representations at the level of context and then sequentially generating the word-, syllable-, and finally letter-level representations (\Cref{fig2}A-B). 

\paragraph{Sequential deactivations of the language hierarchy.}
Although described at the key press level \citep{kornysheva_neural_2019}, the deactivation of neural representations across the language hierarchy has received less attention. We therefore applied the same analysis presented previously at word offset. 
We find that the progressive deactivation of representations depends on their level in the language hierarchy (\Cref{fig2}A). 
Specifically, \emph{after} word offset, the representations of letters go back to chance before those of syllables and words, while context representations are sustained over the longest time period. Overall, this leads to a Spearman correlation between the decay half-times and their linguistic levels of R=0.64$\pm$0.10, $p<10^{-4}$ (\Cref{fig2}B).

\paragraph{Neural representations sustained for longer than their respective execution.}
Consistent with the motor execution literature, the decoding of each representation lasts for a relatively long time period (\Cref{meg raw score sup}) \citep{Averbeck2002, kornysheva_neural_2019, pinet_tracking_2019}. 
For example, each keypress lasts, on average, for 103$\pm$33\,ms, but the corresponding neural representation is decodable from -480\,ms to +640\,ms relative to key press and release, respectively -- \emph{i.e.} 10x longer. 
A similar, although weaker, effect can also be observed for syllables (lasting for 297\,ms, decodable from -480 to +900\,ms, \emph{i.e.} 4x longer) and words (lasting for 799\,ms duration but decodable from -1.5\,s to +1.38\,s, \emph{i.e.} at least 3x longer). 
This phenomenon suggests that at all levels of the language hierarchy, \emph{successive} items are \emph{simultaneously} represented at each instant.

\paragraph{Simultaneous representations of multiple keys during word production.}
To explicitly assess this overlap in terms of letter representations during continuous production, we trained linear classifiers to predict letters at successive positions starting from -- or ending at -- each word (\Cref{fig2}C).
The results show that up to five successive letters can be simultaneously decoded from a single MEG time sample, revealing a superposition of their representations. 
This phenomenon appears to go over word boundaries: he last three keys can be decoded after word offset (\Cref{fig2}D). 
Overall, these analyses confirm that multiple \emph{successive} representations are \emph{simultaneously} present in the brain activity, supporting the representational overlap hypothesis (see \Cref{fig2}C). This result generalizes previous work on serial order processes from two-letter words \citep[e.g.][]{pinet_tracking_2019} to a complex language production setting. 

\begin{figure}[t!]
    \centering
    \includegraphics[width=\linewidth]{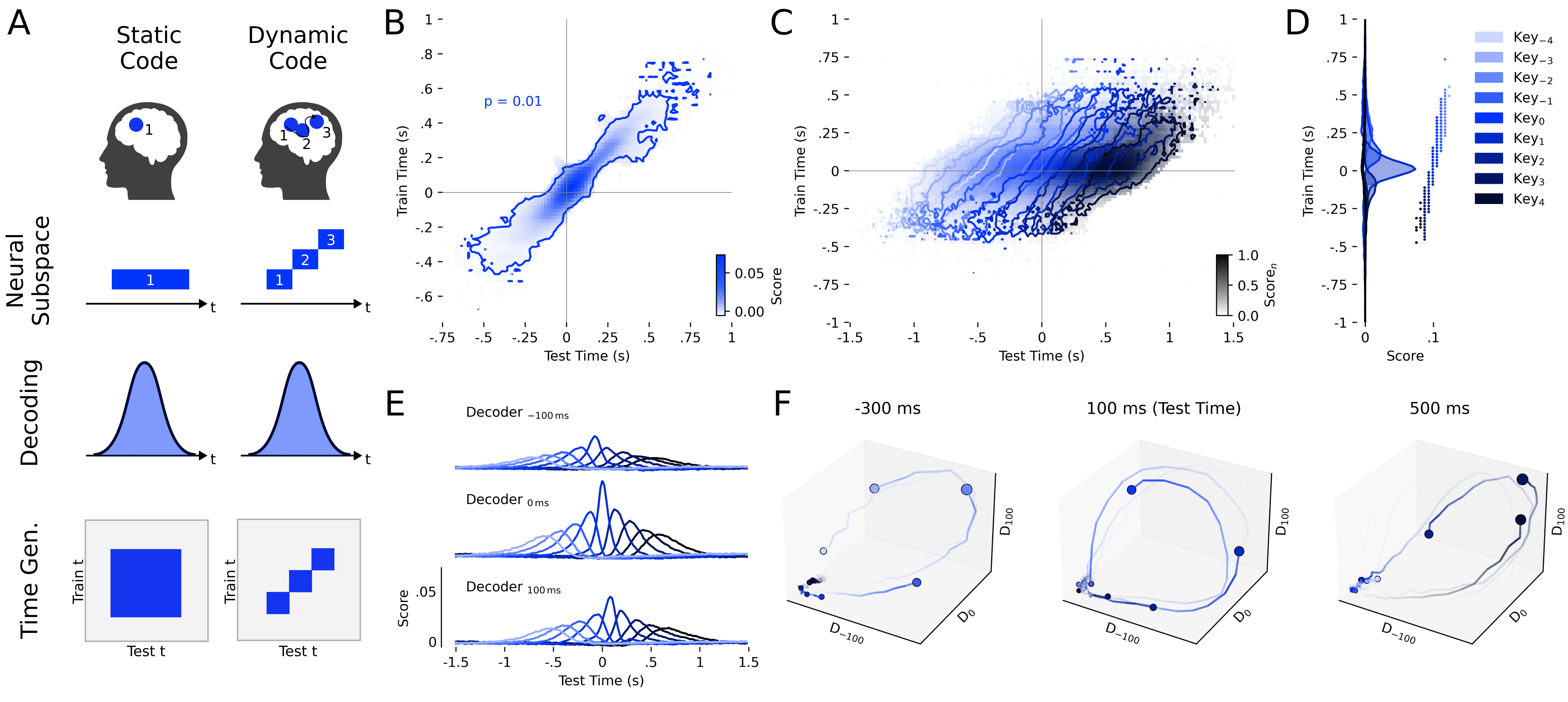}
    \caption{\textbf{Dynamic code of letter representations.} \textbf{A.}  
    Static and dynamic codes may present similar decoding curves (middle) but different temporal generalization matrices (bottom). 
    \textbf{B.} Temporal generalization of letter decoding. The contour indicates $p<0.01$ across subjects. 
    \textbf{C.} Shifted temporal generalization. Each color indicates a different letter relative to the current letter $i$.  
    \textbf{D.} Scores of each decoder trained at each time $t_i$ evaluated on successive letters (same data as the vertical slice of the temporal generalisation matrices at test time = 0\,s in C). Vertical lines indicate statistical significance $p<0.05$. 
    \textbf{E.} Scores of three decoders fitted at -100\,ms, 0\,ms, and 100\,ms on successive letters (i.e. three horizontal slices of the Time Gen. of panel C).
    \textbf{F.} Same as E, plotted as normalized 3D trajectories for three representative test times (-300, 100 and 500\,ms). Each axis represents one decoder. 
}
    \label{fig3}
\end{figure}

\paragraph{Letters are represented with a dynamic neural code. }
How does the brain avoid the interference induced by such overlapping representations? 
One hypothesis is dynamic neural coding. This coding scheme can be viewed as a positional embedding in the context of artificial neural networks \citep{vaswani_attention_2017}. In practice, this consists of "moving" the representation across different neural subspaces. 
Accordingly, the availability of letter information remains constant over time, but \emph{where} this information is represented in neural activity continuously changes (\Cref{fig3}A). 
To test this hypothesis, we implemented temporal generalization analysis \citep{king_characterizing_2014} which consists of evaluating whether the neural representations decoded at a time sample effectively generalize to other time samples. 
If the neural code is stable, then a decoder trained at time $t_i$ should generalize to time $t_j$. 
Conversely, if the neural code is dynamic, then the model trained at $t_j$, but not the model trained at $t_i$, would be able to decode at $t_j$. 
In this view, stable codes lead to square temporal generalization matrices, whereas dynamic codes lead to diagonal matrices (\Cref{fig3}A). 
The results show that letters are represented with a dynamic neural code. The temporal generalization matrix is marked by a diagonal pattern (\Cref{fig3}B). We can then evaluate whether each of these decoders can decode, not only the current letter, but also the preceding and the subsequent letters. The results show that successive letters lead to a sequence of ``parallel diagonals'' (\Cref{fig3}C), with multiple letter representations detectable through decoder generalization at the same time sample (\Cref{fig3}D). We observed a similar, although noisier, result with EEG signals (\Cref{eeg sup key dyna}).

Notably, each of these letter representations moves in a high-dimensional space.
For example, we can focus on three decoders trained at -100\,ms, 0\,ms and +100\,ms, respectively (\Cref{fig3}E), and visualize them as three dimensions of this space (\Cref{fig3}F). 
The decoding scores of successive letters can then be visualized over time as trajectories in this 3D space. At each instant, each letter is captured differentially by the three decoders, as the representations are positioned in different parts of this space. 
Consequently, this dynamic coding scheme allows a linear readout of individual letters without being interfered by the others.



\begin{figure}[t!]
    \centering
    \includegraphics[width=\linewidth]{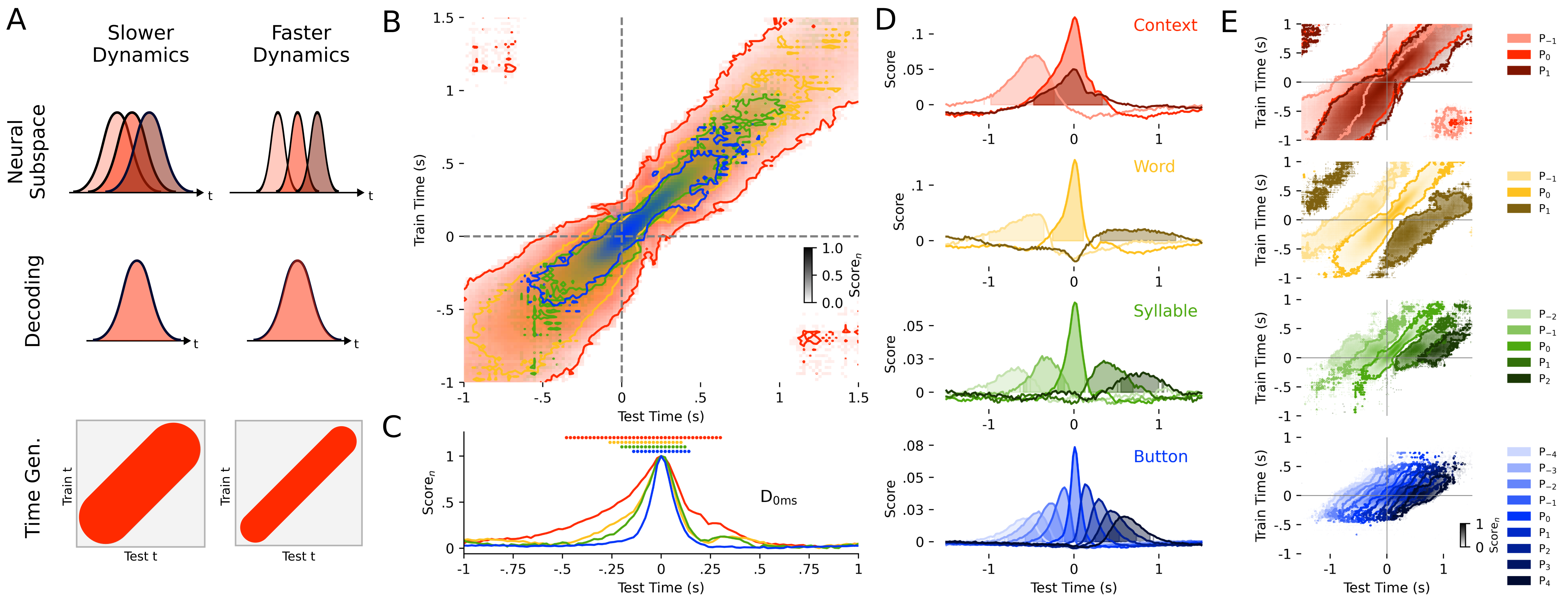}
    \caption{\textbf{Dynamic neural code across the language hierarchy.} \textbf{A.} Hypotheses of slow vs fast neural dynamics can lead to similar decoding curves but different temporal generalization matrices \citep{king_characterizing_2014}. 
    \textbf{B.} Normalized temporal generalization of features time-locked at all feature onset. Contours indicate statistical significance across subjects $p<0.01$. 
    \textbf{C.} Normalized generalization scores of decoders from each linguistic level trained at 0\,ms and tested over all time samples. Horizontal marks indicate statistical significance $p<0.05$. 
    \textbf{D.} Scores of decoder trained at 0\,ms and evaluated across time for each preceding and subsequent feature (i.e. Horizontal slice of panel E). Shaded areas indicate statistical significance $p<0.05$. 
    \textbf{E.} Temporal generalization of previous, current, and next features for each level of the language hierarchy. 
    }
    \label{fig4}
\end{figure}

\paragraph{A hierarchy of dynamical codes. }
Do the different levels of language representations follow the same dynamics (\Cref{fig4}A)?
Here, we hypothesize that the \emph{speed} of the representational dynamics will depend on their level in the language hierarchy. Specifically, context representations would ``move'' more slowly than letter representations. The temporal generalization matrices of contexts should thus be marked by ``wider'' diagonal patterns than those of letters (\Cref{fig4}A). 

All these hypotheses are verified. 
First, each level of representation is marked by a diagonal pattern: \emph{i.e.} the periods during which letters (1.1\,s), syllables (1.4\,s), words (2.8\,s) and contexts (3\,s) are decodable far exceed the periods during which each decoder generalizes (letter : 280\,ms, syllable : 320\,ms, word : 360\,ms, context : 780\,ms for decoder trained at 0\,ms)(\Cref{fig4}B-C). This result confirms that all levels of representation are instantiated by a dynamic neural code.
Second, the speed of these representational dynamics depends on their level in the language hierarchy. The generalization duration of the decoder trained at t=0\,ms, quantified by the difference between the half-times before and after event onset (\Cref{Method}), significantly correlates with the level of the language hierarchy: Spearman R=0.74$\pm$0.06, $p<10^{-5}$. A similar, although noisier, correlation can be observed from EEG signals: Spearman R=0.39$\pm$0.11, $p<10^{-2}$ (\Cref{eeg sup dyna all}).



%
Finally, similarly to letters, these dynamical codes allow successive elements (e.g. syllables, words) to be simultaneously represented in brain activity without sharing the same neural subspaces (\Cref{fig4}D-E).

\section{Discussion}


\paragraph{Results summary.}
To explore the neural mechanisms coordinating language production, we measured the brain activity of 35 skilled typists with either MEG or EEG while they typed sentences on a keyboard. 
We find that four levels of language representations can be linearly decoded from these signals. 
The evolution of these representations follows the predictions of a top-down recruitment of the language hierarchy: before each word is typed, the brain generates context representations, followed by word, syllable, and finally letter representations (\Cref{fig2}B).
While these representations overlap in time (\Cref{fig2}D), we show that they actually exist in separate subspaces of neural activity thanks to a dynamic code (\Cref{fig3}A-C, \Cref{fig4}).

\paragraph{Top-down activation of the language hierarchy.} 
These results confirm the long-standing predictions that language production requires a hierarchical decomposition of sentence meaning into progressively smaller units that ultimately control motor actions \citep{chomsky_aspects_1965,indefrey_spatial_2011,logan_hierarchical_2011}. 
They also extend a growing literature on the temporal characteristics of language representations during production \citep{munding_cortical_2016}. 
For example, MEG results from \citet{carota_time_2022} suggest a progression from semantic to phonological computations before overt articulation in a picture naming task. 
Furthermore, \citet{khanna_single-neuronal_2024} recently showed with single-unit neural recordings that several levels of the language hierarchy are activated following a specific order in the posterior frontal gyrus during speech. 
Our study complements these findings with, here, (i) a complex language production task that reduces sensory confounds, and (ii) a systematic decomposition of the whole-brain activity into a hierarchy of language representations. 
Interestingly, the reported \emph{top-down} sequence of language representations mirrors the \emph{bottom-up} construction of representations observed with fMRI, M/EEG, and intracranial electrophysiology during language perception \citep{millet_toward_2023, li_dissecting_2023, caucheteux_language_2021, goldstein_shared_2022}. However, the question of how closely these perception and production hierarchies share the same neural codes remains open \citep{lin_neural_2022, fairs_shared_2022}. 

\paragraph{A hierarchy of dynamical neural codes.} 
The hierarchy of neural dynamics revealed in this study provides a clear mechanism to simultaneously represent successive features and actions, while avoiding their mutual interference. Our results show that each representation is continuously moving across different subspaces of neural activity -- this specific dynamics thus allows different linear read-outs to target a specific representation in each sequence.  
Critically, we observe that the speed of these dynamics depends on their level in the language hierarchy: at the top level, the representations of context slowly move as the sentence unfolds, whereas at the lowest level, the representations of letters move quickly. 
This specific neural code thus allows the brain to both (i) represent a rich set of linguistic features and (ii) coordinate their chained execution.

\paragraph{Related works.} This dynamic neural code echos recent findings on the neural bases of speech comprehension, which represent \emph{perceived} words and phrases more slowly than those of syllables and phonemes \citep{gwilliams_hierarchical_2024}. Such temporal characterization of representations enriches previous discussions focused on processing stages in word production \citep{munding_cortical_2016}.  
Furthermore, this dynamic neural code has been repeatedly observed in the motor literature. For example, \citet{churchland_neural_2012}, \citet{russo_neural_2020}, and \citet{crowe_dynamic_2014} showed that the neural activity of the primate motor cortex follows dynamic trajectories during the execution of a simple motor task. 
Together, these findings thus suggest that the hierarchical organization of neural dynamics may provide a canonical mechanism to coordinate rapid sequences of representations. 

\paragraph{Anatomical localization.} 
\citet{indefrey_spatial_2011} suggests that word retrieval primarily depends on the middle temporal gyrus (MTG), whereas the preparation of phonological and syllabic representations takes place in posterior temporo-parietal and inferior frontal gyri \citep{hickok_architecture_2014,guenther_neural_2016}. 
While the present M/EEG study is unsuitable to pinpoint the exact brain location of language representations, it complements these seminal anatomical findings by clarifying the temporal unfolding of the neural representations of natural language production. 
Nevertheless, the full spatio-temporal properties of this anatomico-functional hierarchy remains to be fully characterized \citep{price_review_2012, indefrey_spatial_2011, hickok_computational_2012}. 


\paragraph{The modalities of language production.}
The present typing paradigm was primarily motivated to circumvent the artifacts and confounds generated by natural speech.  
However, the architecture of the human brain did not evolve to type on a QWERTY keyboard. 
This design choice thus leaves several questions open. 
First, the present study does not address whether the letter representations correspond to abstract linguistic features or rather to keyboard-specific motor actions. 
Second, whether the neural organization revealed here applies \emph{universally} across the different modalities of human communication remains to be resolved.. 
Nevertheless, beyond its empirical findings, this study provides the analytical and conceptual tools to unravel the neural and computational mechanisms that orchestrate the transformations of thoughts into actions. 

\section{Method} \label{Method}

\subsection{Protocol and preprocessing}
\paragraph{Participants.}
We recruited 35 healthy volunteers at the Basque center on Cognition, Brain and Language (BCBL) (77\% female, age 31.6$\pm$5.2). Twenty joined MEG sessions and twenty joined EEG sessions (five joined both). 
To ensure that the participants could type without looking at the keyboard, we only selected native Spanish speakers who scored more than 80\% accuracy in a dictation typing task, where the keyboard was covered with a cardboard box. 
Each participant provided informed consent for this study and received a 12 euro/hour compensation. The study was approved by the local ethics committee. 
Data from one participant was excluded from the MEG data analysis due to the presence of a metallic component detected during the MEG recording. 

\paragraph{Task.}
Each trial consisted of three phases (\Cref{fig1}A). First, a sentence was presented on the screen through a Rapid Serial Visual Presentation (RSVP) protocol, where words were shown for a random duration between 465-665\,ms. 
Second, participants were tasked to memorize that sentence for a delay period of 1.5\,s, during which only a fixation cross was presented on the screen. 
Finally, participants started typing the sentence from memory once the fixation cross disappeared from the screen. During typing, only a small black square was presented on the screen, rotating 10 degrees clockwise on every keystroke. This helped signal a successful key press without presenting the key input and thus ensured that participants minimized eye movements. Trials were separated by a 1\,s interval. Breaks were introduced every 16 trials. 
Each participant took part in multiple sessions (from 2 to 6), with each session consisting of two blocks of 64 trials. 
To familiarize participants with the protocol, four training trials were performed at the beginning of each block. Data from these trials was excluded from further analyses.
On average, the reading and typing phases lasted for 5.54\,s and 4.70\,s, respectively. 

\paragraph{Stimuli.}
To balance language complexity and repetition, we designed 128 unique declarative sentences, which varied in length (5 - 8 words), meaning, and syntactic structures. All sentences contained a subject, a verb, and an object. The subject and object were a minimal noun phrase formed by a determiner and a noun in singular or plural form. The noun phrase could be supplemented with an adjective (e.g. "the data"/"the general data") or a prepositional phrase (e.g. "the data"/"the data of the transaction"). 
All stimuli were presented in black on a 50\% gray background. During reading, all letters were capitalized and did not contain any diacritic marks. During typing, participants were instructed to type without diacritic marks as if they were typing in upper case. They were also discouraged from using the backspace key. 

\paragraph{Keyboard.}
Standard keyboards contain electronics and metallic parts that lead to major MEG artifacts. To address this issue, we adapted an MR-compatible keyboard from HybridMojo (LLC) by changing the keys supported by a spring (shift, spacebar, backspace) with a custom non-ferromagnetic silver-spring mechanism. During the task, the keyboard was roughly 70\,cm away from the top of the helmet of the MEG scanner.

\paragraph{MEG.}
MEG recordings were acquired with a 306 Megin system (102 magnetometers and 204 planar gradiometers) at a sampling rate of 1\,kHz, with an online high-pass filter set at 0.1\,Hz and a low-pass filter at 330\,Hz. Electrooculograms (EOG) and electrocardiograms (ECG) were recorded through auxiliary channels. Five head position index (HPI) coils were used for tracking head position. The location of each coil relative to the anatomical fiducials (nasion, left and right preauricular points) was defined with a 3D digitizer (Fastrak Polhemus, Colchester, US). HPI were acquired at the beginning of each session. MEG signals were bandpass filtered between 0.5\,Hz and 45\,Hz with MNE-python's default parameters and downsampled to 50\,Hz \citep{gramfort_meg_2013}. For \Cref{fig1}C-D, we plot the baseline-corrected (-250\,ms-0\,ms), maxfiltered signal-space-projected (SSP) gradiometers median-averaged across subjects, except for the wait panel where no baseline correction is applied to vizualize the activity during this period.  

\paragraph{EEG.}
EEG recordings were acquired with an actiCAP slim system from BrainVision at a sampling rate of 1\,kHz, with an online high-pass filter set at 0.02\,Hz. A total of 61 EEG channels were recorded together with 3 ocular channels. We applied the same preprocessing methods as described in the MEG section on EEG signals. 

\paragraph{Events construction.}
Language can be described at different levels of representations (letters, syllables, words, sentences).
We defined an event’s onset as the press of its first key and its offset as the release of its last key (i.e. syllable event "con" starts when "c" is pressed and ends when "n" is released, \Cref{fig1}B). 
To omit typographical errors from the typed sentences, we computed the Levenshtein distance between the true and typed sentences and kept the subset of keystrokes that matched the true sentence.
Boundaries between typed words are given by space keys while boundaries between syllables are defined by manual annotation. Space keys and mistyped syllables are excluded from subsequent analyses. Epochs are time-locked either to the onset of words or the onset of all features in the language hierarchy. 

\subsection{Language representations}
We featurized each level of the language hierarchy into target vector representations (\Cref{fig1}B). All features are generated from text strings free from diacritic marks. 

\paragraph{Context representations: GPT2.}
We used GPT2's hidden activations \citep{radford_language_2019} to represent the past context of each word. Specifically, for a target $i^{th}$ word in the sentence, we tokenized the text up to and including the current word (0-$i^{th}$ word) with the GPT2 tokenizer and input this context to an inference-only pre-trained GPT2 model as provided by HuggingFace. We defined the context representation as the activation of the $i^{th}$ word in the $8^{th}$ transformer layer (d=768), similar to \citet{caucheteux_brains_2022}. If a word is represented by multiple tokens, we averaged the activations corresponding to the tokens of the target word. 

\paragraph{Word representations: Spacy.}
To generate word-level representations, we used a pre-trained Spanish language model (es\_core\_news\_lg) from the Spacy library \citep{ines_montani_explosionspacy_2023}, leading to a $d=300$ embedding of each word. 

\paragraph{Syllable representations: FastText.}
To obtain vectorized representations of sublexical strings, we used FastText -- a subword embedding model introduced by \citet{bojanowski_enriching_2017}.
%
We trained a custom FastText model using the Spanish wikidump corpus (date July 19th, 2024)
 \citep{joulin_fasttextzip_2016, wikidumps_wikimedia_nodate}. The corpus was cleaned and preprocessed using the wp2txt pipeline \citep{hasebe_wp2txt_2023}. The model was trained with default hyperparameters, except for n-gram length which ranged between 1 to 5 and a dimension size set to $d=300$. Each syllable was then extracted from the library of n-gram vectors. 

\paragraph{Letter representations: One-Hot-Encoder.}
For the letter representations, we simply use a one-hot-embedding of the letter identities. 

\subsection{Decoding analyses}

We deployed four variants of decoding analyses, each addressing a distinct question about the temporal evolution of neural representations. 

\paragraph{Time decoding.}
To assess when a feature is represented, we conducted decoding analyses over time relative to each event. Given $X\in \mathbb{R}^{w\times s\times t}$, with $s$ being the number of M/EEG sensors time-locked to the onset or the offset of $w$ words over $t$ time samples, and their corresponding target vectors $y$, we fitted and evaluated a linear decoder at each time sample $t$. 

\paragraph{Shifted time decoding.} 
The above analyses decode $y_i$ time-locked to event $i$. To evaluate the overlap of successive representations, we can extend these analyses to decode $y_j$ time-locked to event $i$. 
Concretely, we trained and evaluated a series of decoders to predict $y_j$ relative to $t_i$. 

\paragraph{Temporal generalization.} To assess the stability of neural representations, we conducted temporal generalization analyses \citep{king_characterizing_2014}. This consists of training a decoder $d_{t\rightarrow t'}$ on neural signals $X$ at time $t$ and testing it on $t'$. If a neural representation is stable, then $d_{t\rightarrow t'}$ should lead to decoding performances similar to $d_{t\rightarrow t}$. By contrast, if the neural code is dynamic, meaning that the neural representation of a feature changes over time, then $d_{t\rightarrow t'}$ should perform worse than $d_{t\rightarrow t}$. 

\paragraph{Shifted temporal generalization.}
Temporal generalization can be further extended to assess the decoder's generalization of successive features. For this, we evaluated whether each time-decoder (trained with $y_i$ at $t_i$) effectively decodes, on a test set, the preceding or subsequent features $y_j$. 

\paragraph{Decoding models.} 
We used a ridge linear regression (\texttt{RidgeCV}) as implemented in scikit-learn \citep{pedregosa_scikit-learn_2011}, with an alpha regularization (alpha\_per\_target=\text{True}) spanning 50 log-spaced values between $10^{-1}$ and $10^{10}$ with  default parameters. 
Neural signals are standardized (zero-meaned and scaled to unit variance) per channel before model fitting. 

\paragraph{Train-test procedure.} 
We trained and tested each decoder in a within-subject fashion. We used a group k-fold (K=5) cross-validation to assess decoding performance, where the group of each feature is assigned based on unique sentences. Groups are split into a train set (80\%) and a test set (20\%). 

\paragraph{Evaluation metrics.} 
Unless stated otherwise, we evaluated decoding performance with a Pearson correlation between the true and the decoded features averaged across $d$. For letter decoding, we report the difference in the balanced accuracy score, which is the average of recall measured in each class, between a true model and a dummy model that predicts the most frequent class label. Each representation of language does not have the same space or dimensionality. Consequently, we compared the relative evolution of decoding performances $Score_n$ with a ``normalized decoding performance'': \emph{i.e.} $Score_n=\frac{Score-min(Score)}{max(Score)-min(Score)}$, where $min$ and $max$ are applied over time.


\paragraph{Half-times.}
To assess when each representation rise relative to event onset and offset, we computed the `half-time' of decoding scores, defined as the moment when decoding performance reaches 50\% of the maximum decoding score within the window of interest per participant. We focused on the [-1, 0]\,s window prior to feature onset and [0, 1]\,s after feature offset. 

\paragraph{Significance testing.} To test whether the decoding scores are significantly different from chance, we used the Wilcoxon signed rank test implemented in the Scipy library \citep{virtanen_scipy_2020} across subjects. In the case of multiple comparisons, we further adjusted the p-values with a False Detection Rate (FDR) as implemented in MNE \citep{gramfort_meg_2013}. 

\paragraph{Companion paper.} We explore the possibility to efficiently decode text from these brain signals in a companion paper \citep{levy_brain--text_2025}. Note that the A and D panels of Fig. \ref{fig1} are shared between these two studies.

\section{Acknowledgements}
The authors would like to thank Maite Kaltzakorta, Manex Lete, Jessi Jacobsen, Daniel Nieto, Jone Iraeta, Araitz Garnika, Jaione Bengoetxea, Natalia Louleli, Naroa Miralles, Eñaut Zeberio, Craig Richter, Amets Esnal, and Olatz Andonegui. 
This research is supported by the Basque Government through the BERC 2022-2025 program and Funded by the Spanish State Research Agency through BCBL Severo Ochoa excellence accreditation CEX2020-001010/AEI/10.13039/501100011033.
FXA was supported within the Institute of Convergence ILCB by grants from France 2030 (ANR-16-CONV-0002) and the Excellence Initiative of Aix-Marseille University (A*MIDEX).
Parts of this research were carried within the European Union's Horizon 2020 research and innovation programme under the Marie Skłodowska-Curie grant agreement No 945304 - Cofund AI4theSciences hosted by PSL University.

\clearpage
\newpage
\bibliographystyle{assets/plainnat}
\bibliography{main}

\newpage
\section{Supplementary Information} \label{Supplementary}
\renewcommand{\thefigure}{S\arabic{figure}}
\setcounter{figure}{0}

\begin{figure}[h]
    \centering
    \includegraphics[width=\linewidth]{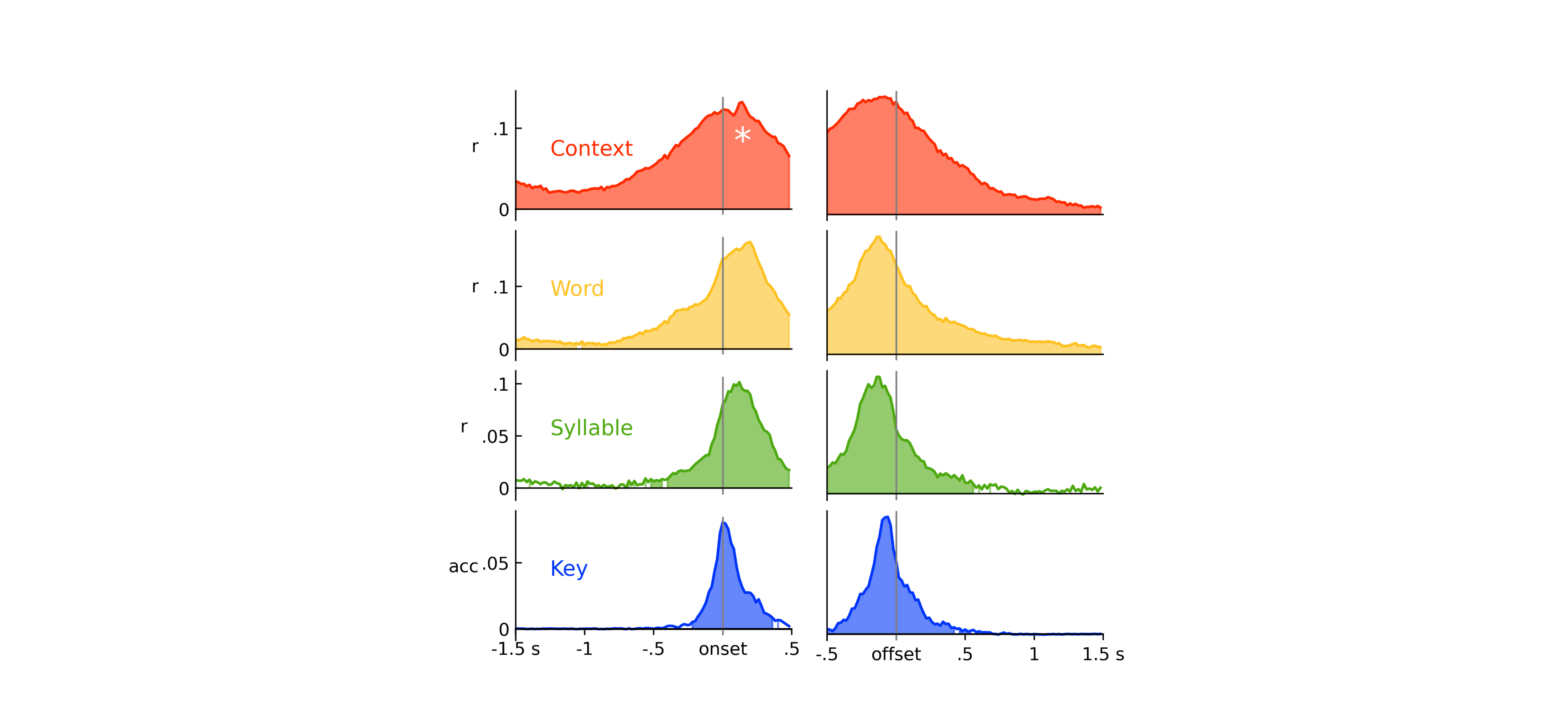}
    \caption{MEG raw decoding score of the language hierarchy in \Cref{fig2}B. Left column time-locked on word onset and right column to word offset. Shaded area indicate significant decoding score $p<0.05$. r stands for Pearson correlation and acc stands for balanced accuracy.}
    \label{meg raw score sup}
\end{figure}

\begin{figure}[h]
    \centering
    \includegraphics[width=\linewidth]{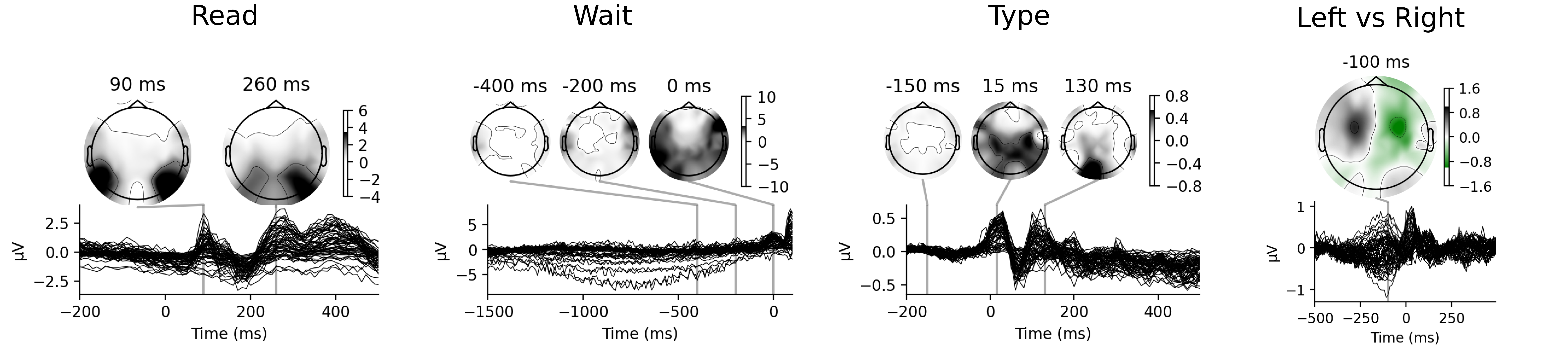}
    \caption{Evoked response of EEG for three phases of the typing tasking (same as \Cref{fig1}C-D). Type response plotted with baseline correction between -250\,ms and -0.05\,ms. No baseline correction applied otherwise. }
    \label{eeg topo}
\end{figure}

\begin{figure}[h]
    \centering
    \includegraphics[width=\linewidth]{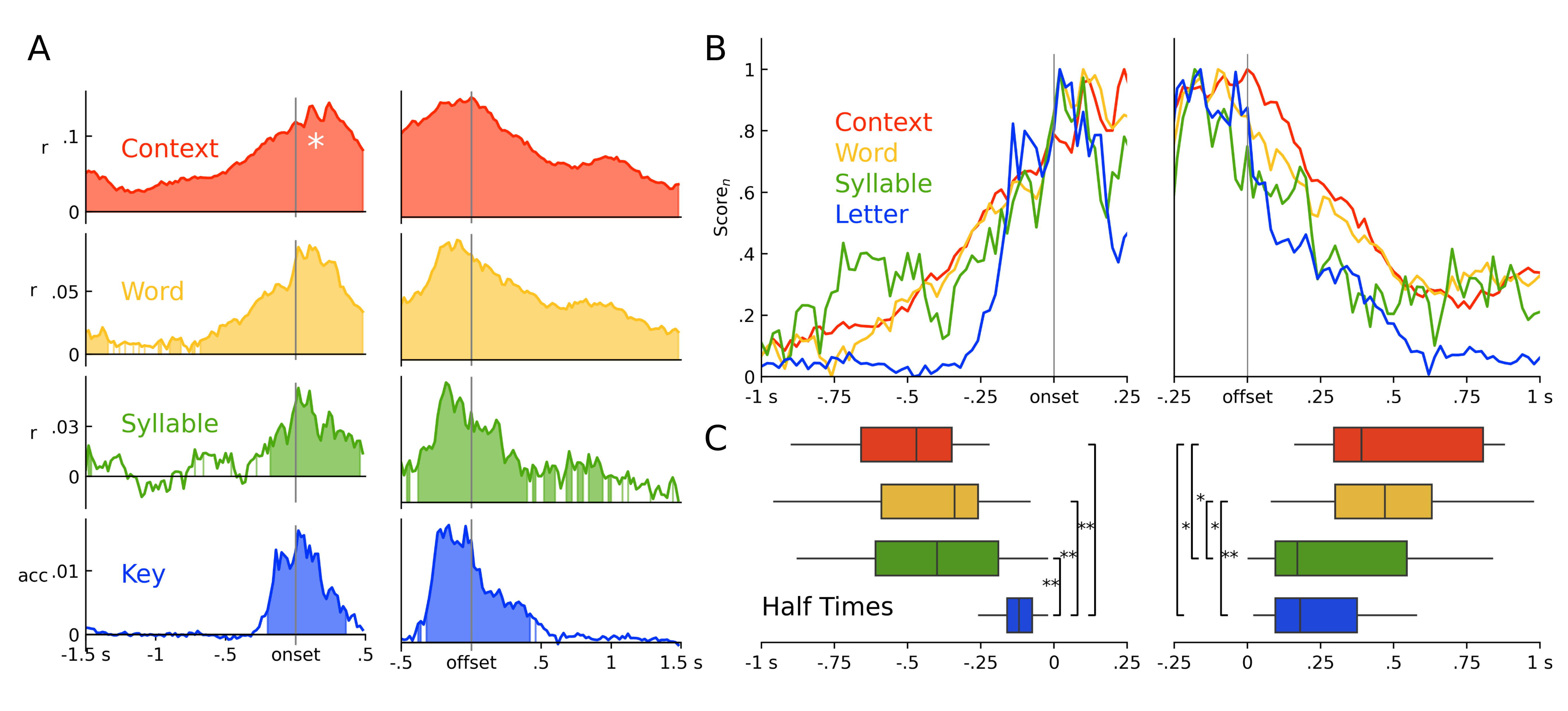}
    \caption{Decoding of the language hierarchy with EEG, replication of \Cref{fig2}A-B. \textbf{A.} Raw score. \textbf{B.} Normalized score. Significant decoding starts at -1.5\,s for contexts, -1.5\,s for words, -180\,ms for syllables, and -200\,ms for letters before word onset. After word offset, contexts can be decoded till 1.46\,s, words till 1.46\,s, syllables till 940\,ms, and letters till 420\,ms.  \textbf{C.} Half time across subjects. Significant pair-wise comparisons in half-time are indicated by brackets between feature levels (*: $p<0.05$, **:$p<0.01$). Significant comparisons among all pair-wise tests are shown unless adjacent levels are consistently different across the hierarchy. Peak decoding scores at word onset: context 0.14$\pm$0.01, word 0.09$\pm$0.01, syllable 0.04$\pm$0.01, letter 1.6\%$\pm$0.2\%. }
    \label{eeg sup hierar}
\end{figure}

\begin{figure}[h]
    \centering
    \includegraphics[width=\linewidth]{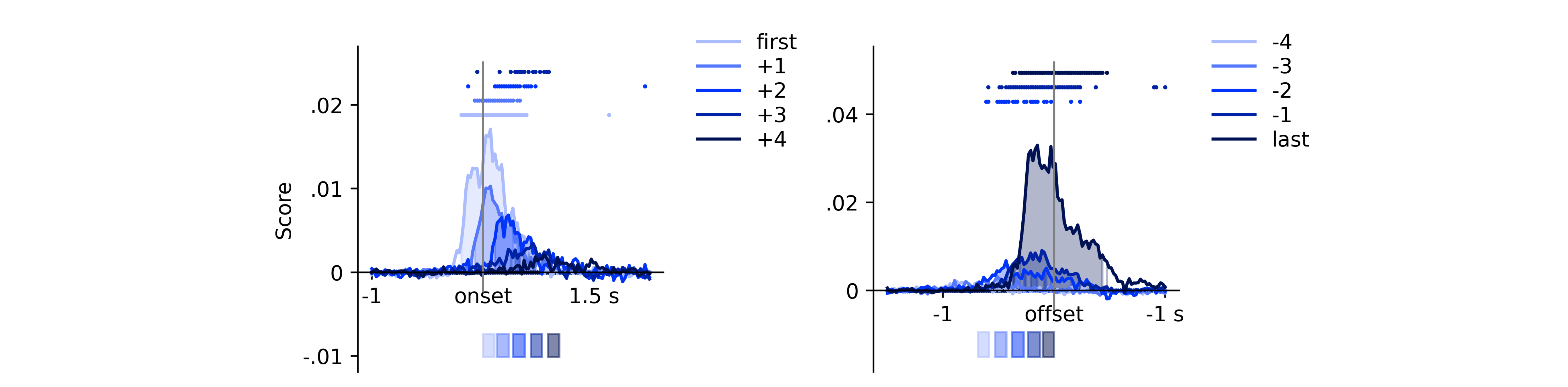}
    \caption{EEG replication of \Cref{fig2}D. Scattered dots and shading indicate time steps with significant decoding scores ($p<0.05$). Rectangles below the x-axis indicate the median start and duration of a key at positions $i$ relative to the first or last keys of the words. }
    \label{eeg sup successive keys}
\end{figure}

\begin{figure}[h]
    \centering
    \includegraphics[width=\linewidth]{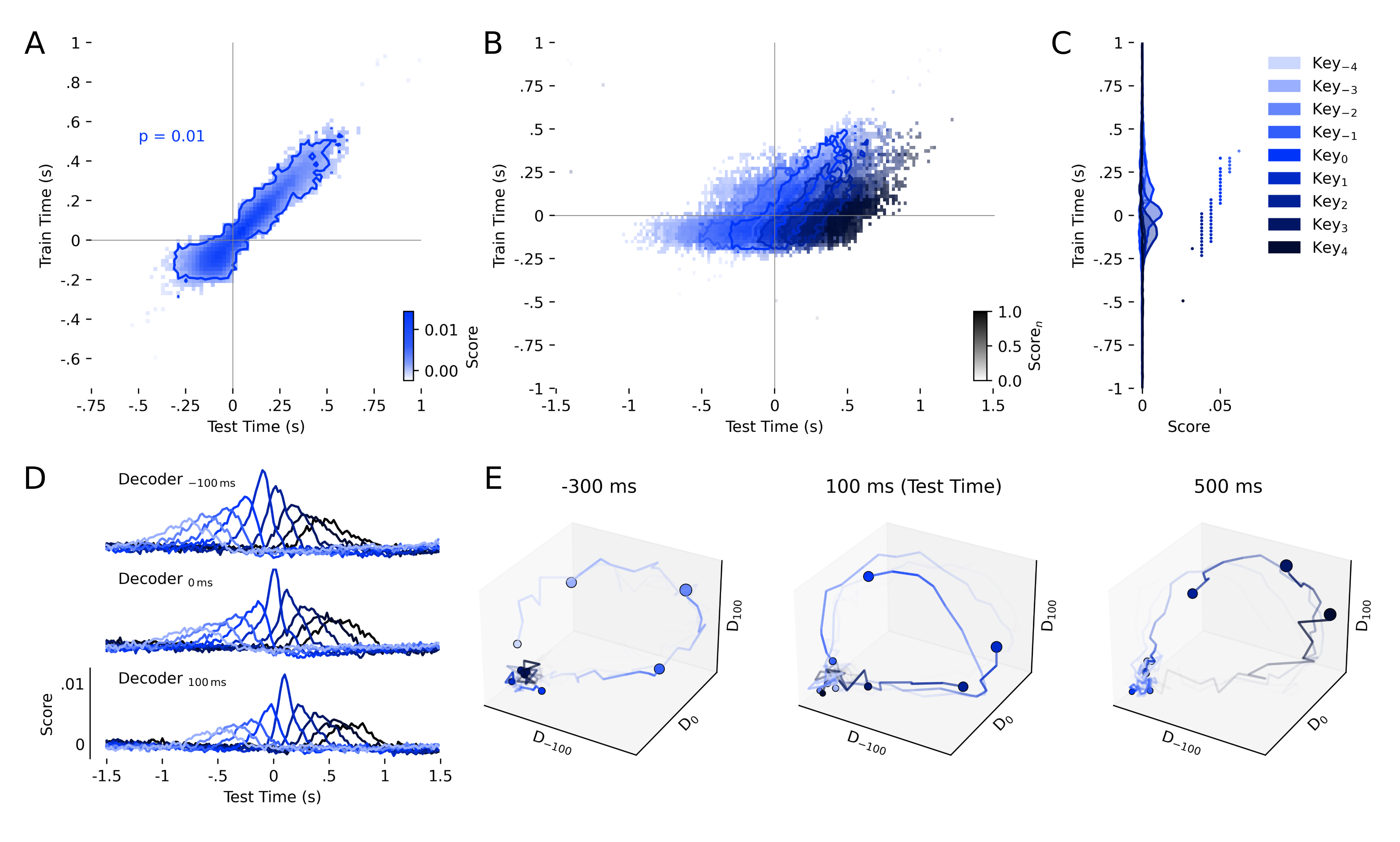}
    \caption{EEG replication of \Cref{fig3}B-F.  
    \textbf{A.} Temporal generalization of letter decoding. The contour indicates $p<0.01$ across subjects. 
    \textbf{B.} Shifted temporal generalization.  
    \textbf{C.} Scores of each decoder trained at each time $t_i$ evaluated on successive letters (same data as the vertical slice of the temporal generalisation matrices at test time = 0\,s in C). Vertical lines indicate statistical significance $p<0.05$. 
    \textbf{D.} Scores of three decoders fitted at -100\,ms, 0\,ms, and 100\,ms on successive letters (i.e. three horizontal slices of the Time Gen. of panel B).
    \textbf{E.} Same as D, plotted as normalized 3D trajectories for three representative test times (-300, 100 and 500\,ms). Each axis represents one decoder. 
    }
    \label{eeg sup key dyna}
\end{figure}

\begin{figure}
    \centering
    \includegraphics[width=\linewidth]{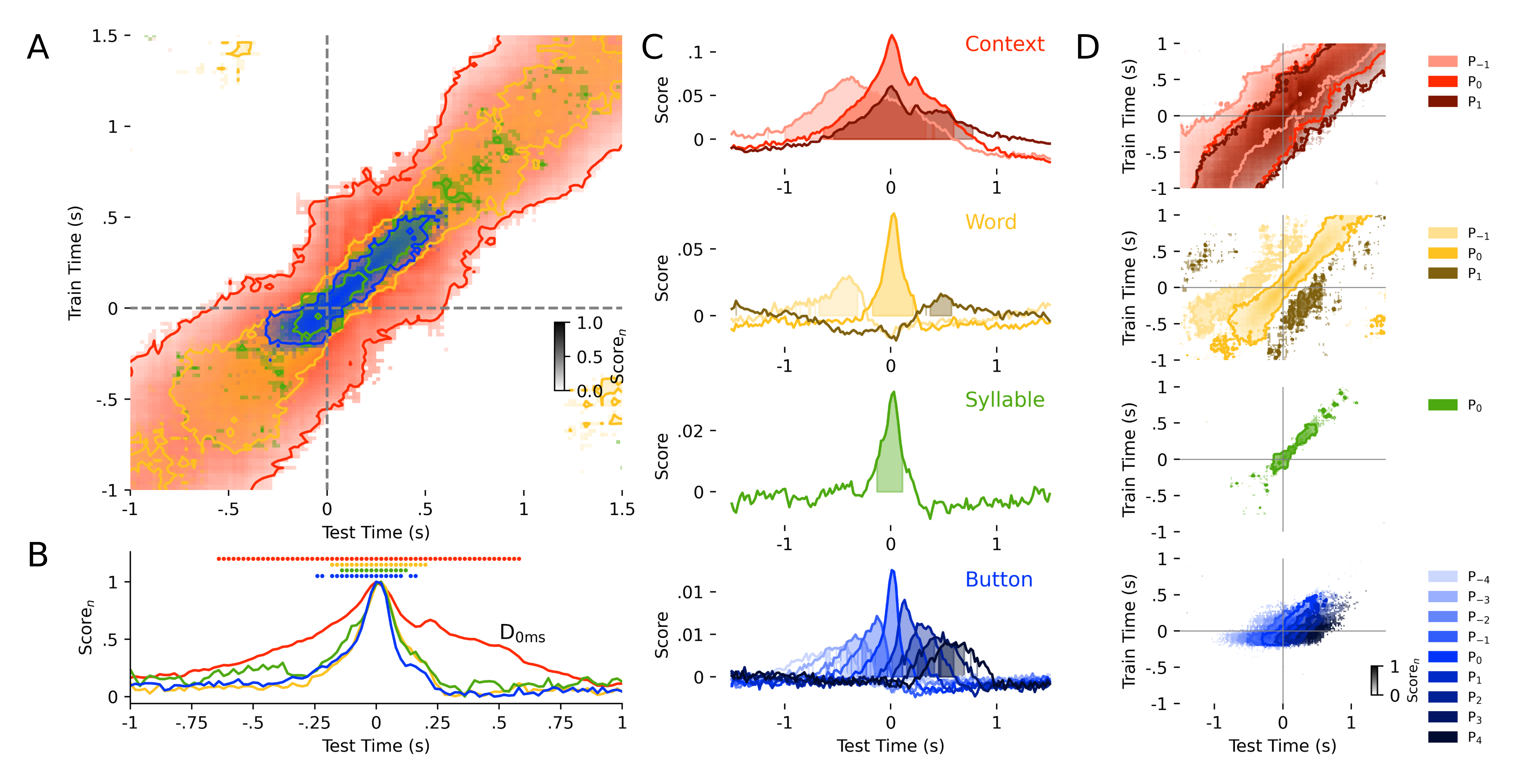}
    \caption{EEG replication of \Cref{fig4}B-E. 
    \textbf{A.} Normalized temporal generalization of features time-locked at all feature onset. Contours indicate statistical significance across subjects $p<0.01$. 
    \textbf{B.} Normalized generalization scores of decoders from each linguistic level trained at 0\,ms and tested over all time samples. Horizontal marks indicate statistical significance $p<0.05$. 
    \textbf{C.} Scores of decoder trained at 0\,ms and evaluated across time for each preceding and subsequent feature (i.e. Horizontal slice of panel D). Shaded areas indicate statistical significance $p<0.05$. 
    \textbf{D.} Temporal generalization of previous, current, and next features for each level of the language hierarchy. }
    \label{eeg sup dyna all}
\end{figure}

\end{document}